\documentclass[fp]{jpsj3}
\usepackage{txfonts}
\usepackage{xcolor}

\usepackage[dvipdfmx]{graphicx}

\title{Machine learning forces trained by Gaussian process in liquid states:
Transferability to temperature and pressure}

\author{Ryo Tamura$^{1,2,3,4}$\thanks{TAMURA.Ryo@nims.go.jp}, 
Jianbo Lin$^1$, 
and Tsuyoshi Miyazaki$^{1,2,5}$\thanks{MIYAZAKI.Tsuyoshi@nims.go.jp}}
\inst{$^1$International Center for Materials Nanoarchitectonics (WPI-MANA),
National Institute for Materials Science, 1-1 Namiki, Tsukuba, Ibaraki 305-0044, Japan \\
$^2$Research and Services Division of Materials Data and Integrated System,
National Institute for Materials Science, 1-2-1 Sengen, Tsukuba, Ibaraki 305-0047, Japan \\
$^3$Graduate School of Frontier Sciences, 
The University of Tokyo, 5-1-5 Kashiwanoha, Kashiwa, Chiba 277-8568, Japan\\
$^4$Center for Advanced Intelligence Project, RIKEN, 1-4-1 Nihombashi, Chuo-ku, Tokyo, 103-0027, Japan \\
$^5$Department of Physics \& Astronomy, University College London, Gower Street, London WC1E 6BT, U.K.} %\\

\abst{
We study a generalization performance of the machine learning (ML) model to predict the atomic forces within the density functional theory (DFT). 
The targets are the Si and Ge single component systems in the liquid state. 
To train the machine learning model, Gaussian process regression is performed with the atomic fingerprints which express the local structure around the target atom. 
The training and test data are generated by the molecular dynamics (MD) based on DFT. We first report the accuracy of ML forces when both test and training data are generated from the DFT-MD simulations at a same temperature. 
By comparing the accuracy of ML forces at various temperatures, it is found that the accuracy  becomes the lowest around the phase boundary between the solid and the liquid states. 
Furthermore, we investigate the transferability of ML models trained in the liquid state to temperature and pressure. 
We demonstrate that, if the training is performed at a high temperature and if the volume change is not so large, the transferability of ML forces in the liquid state is high enough, while its transferability to the solid state is very low. 
}

\kword{Atomic force, Gaussian process regression, Density functional theory, Molecular dynamics, Liquid state}

\begin{document}
\maketitle

%%%%%%%%%%%%%%%%%%
\section{Introduction}
%%%%%%%%%%%%%%%%%%

%General
Data-driven techniques are becoming more and more valuable in computational materials science\cite{Rajan-2005,Oganov-2006,Hautier-2010,Rupp-2012,Snyder-2012,Montavon-2013,Pilania-2013,Meredig-2014,Seko-2015,Tamura-2017,Hegde-2017,Ju-2017,Ramprasad-2017,Brockherde-2017,Yamashita-2018,Tamura-2018,Sumita-2018}.
One of the important issues in this trend is to predict atomic forces by machine learning (ML) technique, and we refer to the atomic forces by ML as ML forces in this paper. Classical molecular dynamics (MD) simulations using empirical force fields have been playing important roles to understand various phenomena of materials at atomic scale, but the reliability of the force fields is a problem in many cases. On the other hand, density functional theory (DFT) calculations can provide reliable atomic forces even if the experimental information is limited. However, since the computational cost of the DFT calculations is much more expensive than classical force fields especially for large systems, both system size and simulation time of DFT-MD simulations are limited. If we can develop ML forces having almost the same accuracy of DFT, the cost of the force calculations would be much cheaper and we can 
perform long-time MD simulations of large systems with DFT level accuracy.
We expect this would realize an acceleration of materials developments led by computational materials science.

% Force

\begin{figure}[b]
\begin{center}
\includegraphics[trim=0mm 0mm 0mm 0mm ,scale=1.0,angle=0]{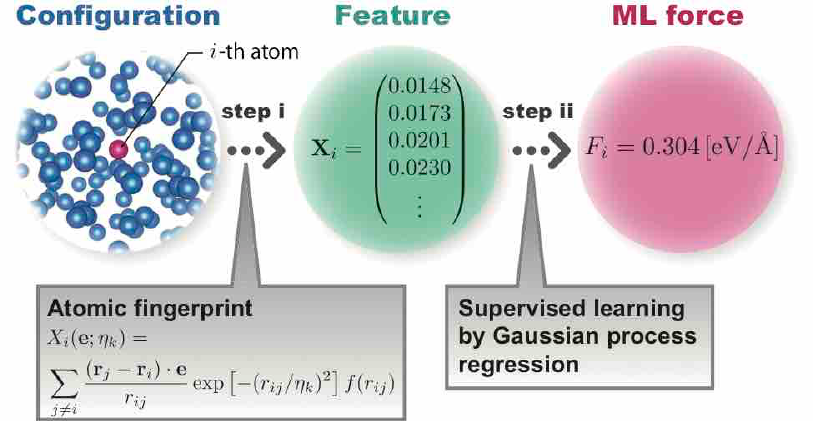} 
\end{center}
\caption{\label{fig:intro}
(Color online) Steps to train machine learning forces.
}
\end{figure}

Strategy to train ML model to predict atomic forces is basically categorized into two groups.
First one is to train the total energy (potential energy surface) calculated by DFT.
In this case, we usually assume the total energy by ML is expressed as the sum of one-body, two-body,
and many-body terms, or the sum of the energy of each atom.
Then, the atomic forces are obtained by the derivative of ML potential with respect to the atomic positions\cite{Behler-2007,Behler-2011,Behler-2011a,Bartok-2013,Pham-2016,Bartok-2017,Deringer-2017,Chen-2017,Li-2017,Dral-2017,Kobayashi-2017}.
The advantage of this method is that the potential and atomic forces can be obtained simultaneously by one ML model.
The other scheme is to directly train the DFT atomic forces\cite{Botu-2015,Li-2015,Botu-2017,Botu-2017a,Huan-2017}.
In this work, we use this method since we expect the accuracy of ML forces would be better. 
It should be noted that, even though the total energy is not calculated with this method, the calculation of energy difference, such as free energy profile, is possible by the thermodynamic integration or blue moon ensemble methods. Obviously, MD simulations are also possible.
In this method, 
atomic forces are predicted by following two steps (Fig.~\ref{fig:intro}).
(i)  The local structure of atomic configurations around the target atom is converted to a feature vector, such as 
atomic fingerprint suggested by Botu and Ramprasad\cite{Botu-2015}.
Since this step is not negligible to obtain highly accurate ML forces,
attempts to develop new methods have been continued.
(ii) Prediction is performed using ML model which is trained by supervised learning methods when the label is atomic forces calculated by DFT calculations.
In general, 
regression techniques such as linear regression, neural-network regression, and Gaussian process regression are used to build ML models.
A lot of demonstrations for various systems including amorphous and multicomponent systems have been reported,
and there are many examples showing that such ML forces have higher accuracy than classical force fields, and high accurate MD simulations can be performed.

For practical use of a ML model, its generalization performance is also important. 
In the actual research, we often need to treat irregular and aperiodic systems having the defects, surfaces, interfaces, locally stressed or heated regions. 
However, it is often difficult and expensive to increase the training data by DFT calculations. 
If possible, we should develop a trained ML model which can correctly predict atomic forces under various conditions such as different temperatures and/or pressures.
To address this challenge, we previously investigated the transferability of ML models for atomic forces in the solid state\cite{Suzuki-2017}.
We showed that the ML model trained at a high temperature in the solid state can predict atomic forces of the solid state in a wide range of temperatures with high accuracy.
However, since generalization performance for more complicated systems is still an open question, the potential of the ML forces is yet to be understood.

%Motivation

In this paper,
we focus on the ML forces for liquid state.
We study the generalization performance of ML models to predict atomic forces when the training data are sampled in the liquid states of the Si or Ge single component systems.
Here, the atomic fingerprint suggested by Botu and Ramprasad is used as the feature of atomic configurations, and Gaussian process regression is performed as the ML method.
In particular,
we address the following three issues:
(i) Accuracy of the ML forces in the liquid state.
(ii) Transferability of ML models to various temperatures.
(iii) Transferability of ML models to various pressures.

%Organization
The organization of this paper is as follows.
Section~\ref{sec:methods} explains the methods to train ML forces,
that is, the definition of atomic fingerprint and Gaussian process regression.
Details of DFT-MD simulations which generate training data and test data are also provided.
In Sec.~\ref{sec:forces},
we show the properties of forces obtained by the DFT-MD simulations of a Si or Ge single component system in the solid and liquid phases. 
In Sec.~\ref{sec:acurracy},
the accuracy of ML forces trained at each temperature is evaluated.
We report a strange behavior that the accuracy of the ML forces around the phase boundary between the solid and liquid is lower than those at other temperatures.
We clarify that the atomic fingerprint used in this work is not capable to capture the middle-range behavior which is more important for low temperature liquid state, 
and this is the reason of strange behavior of ML forces.
Sections~\ref{sec:trans_temp} and \ref{sec:trans_pre} investigate the transferability of ML models at various temperatures and pressures, respectively. 
In the liquid state, we show that the transferability of the ML models to temperature and to pressure is satisfactory, if the volume change is not very large.
On the other hand,
it is concluded that the ML model trained in the liquid state cannot be used in the solid state.
It is because there is no training data generated in the liquid state which is similar to the target test data in the solid state in atomic fingerprint space.
Section~\ref{sec:summary} is the discussion and summary.

%%%%%%%%%%%%%%%%%%
\section{Methods} \label{sec:methods}
%%%%%%%%%%%%%%%%%%

\subsection{Atomic fingerprint for atomic forces}

To train a ML model to predict atomic forces,
we use an atomic fingerprint which was firstly developed by Botu and Ramprasad\cite{Botu-2015}.
The atomic fingerprint expresses a local structure around the target atom and is useful to directly treat the atomic forces in ML.
Note that it is closely related to the radial term of symmetry functions in the Behler-Parrinello method\cite{Behler-2007}.
Since atomic forces are three dimensional vectors,
atomic forces along a specific direction such as $x$-component in Cartesian coordinates are generally trained.
In this work, we train a force component along a randomly selected direction, whose unit vector is $\mathbf{e}$, for each training data. 
The force component and the atomic fingerprint are expressed as
\begin{align}
F_i^u (\mathbf{e}) &= \mathbf{F}_i^u \cdot \mathbf{e}, \\
X_i^u (\mathbf{e}; \eta_k) &= \sum_{j \neq i} \frac{(\mathbf{r}_j^u - \mathbf{r}_i^u) \cdot \mathbf{e}}{r_{ij}^u} \exp \left[- (r_{ij}^u/\eta_k)^2 \right] f(r_{ij}^u). \label{eq:fingerprint}
\end{align}
Here, $\mathbf{F}_i^u$ and $\mathbf{r}_i^u$ are the atomic force and position of the $i$th atom in the $u$th configuration, respectively.
In Eq. ~(\ref{eq:fingerprint}), 
$r_{ij}^u = |\mathbf{r}_j^u - \mathbf{r}_i^u|$ is the distance between the atom $i$ and its neighbor atom $j$, and $\eta_k \, (k=1,...,K)$ is a decay rate for this distance.
A cutoff function $f(r_{ij}^u)$ is given by
\begin{eqnarray}
f(r_{ij}^u) = 
\begin{cases}
0.5 \left[ \cos \left(\pi r_{ij}^u /R_{\rm c} \right) + 1 \right] & {\rm for} \  r_{ij}^u \le R_{\rm c} \\
0 & {\rm for}  \ r_{ij}^u > R_{\rm c}
\end{cases},
\end{eqnarray}
where $R_{\rm c}$ is a cutoff radius.
By considering $K$ types of decay rate,
$K$-dimensional fingerprint vector corresponding to a feature in ML is obtained as 
\begin{eqnarray}
\mathbf{X}_i^u (\mathbf{e}) = \left(X_i^u (\mathbf{e}; \eta_1),...,X_i^u (\mathbf{e}; \eta_K) \right)^\top.
\end{eqnarray}

\subsection{Gaussian process regression}

Gaussian process regression (GPR) is one of the supervised learning methods where each training data has the feature (e.g., fingerprint vector) and the label (e.g., atomic force component).
GPR predicts a value of label at any feature vector by non-linear functions\cite{Bishop-2006}.
We train GPR using the Bayesian optimization library: COMBO\cite{Ueno-2016}.
In this library, Gaussian process is approximated by Bayesian linear model with a random feature map\cite{Rahimi-2007}.
The hyperparameters are automatically determined by maximizing the type-II likelihood\cite{Rasmussen-2006},
and overfitting is prevented by regularization even if the dimension of inputted features is high.
The advantage to use COMBO is that the computational time scales as a linear function against the number of training data points.
Notice that to train GPR, components in fingerprint vectors are normalized by using z-score.

\subsection{DFT-based MD simulations}

To generate the data sets of atomic forces, we perform DFT-MD simulations with a linear-scaling method using the CONQUEST code\cite{CONQUEST}. The system contains 1000 atoms in a cubic cell, whose side length is 27.15 ${\rm \AA}$ and 28.28 ${\rm \AA}$ for Si and Ge cases, respectively. 
The details of the linear-scaling DFT-MD method are explained elsewhere\cite{Bowler-2002,Miyazaki-2004,Bowler-2006}.

For the calculation conditions, we employ the local density approximation (LDA) with the standard Ceperley-Alder exchange-correlation functional. Troullier-Martins type norm-conserving pseudopotentials and the pseudo-atomic orbital (PAO) basis sets are generated by Siesta code\cite{Soler-2002}. We use a minimal basis set, whose accuracy was reported in Ref.\cite{Miyazaki-2007}, and the cutoff energy for the charge density grid is 80 Hartree. 
Density matrix minimization (DMM) method is performed to realize a linear-scaling DFT-MD simulations\cite{Arita-2014}. Cutoff range of the auxiliary density matrix ($L$-matrix) in the DMM method is 16.0 bohr for both Si and Ge cases.  
Using the Nose-Hoover chain thermostats, constant temperature (NVT) simulations\cite{Hirakawa-2017} are conducted with a time step of 1 femto second (fs).

%%%%%%%%%%%%%%%%%%
\section{Atomic forces by DFT-MD simulations} \label{sec:forces}
%%%%%%%%%%%%%%%%%%

\begin{figure*}[t]
\begin{center}
\includegraphics[trim=0mm 0mm 0mm 0mm ,scale=0.8,angle=0]{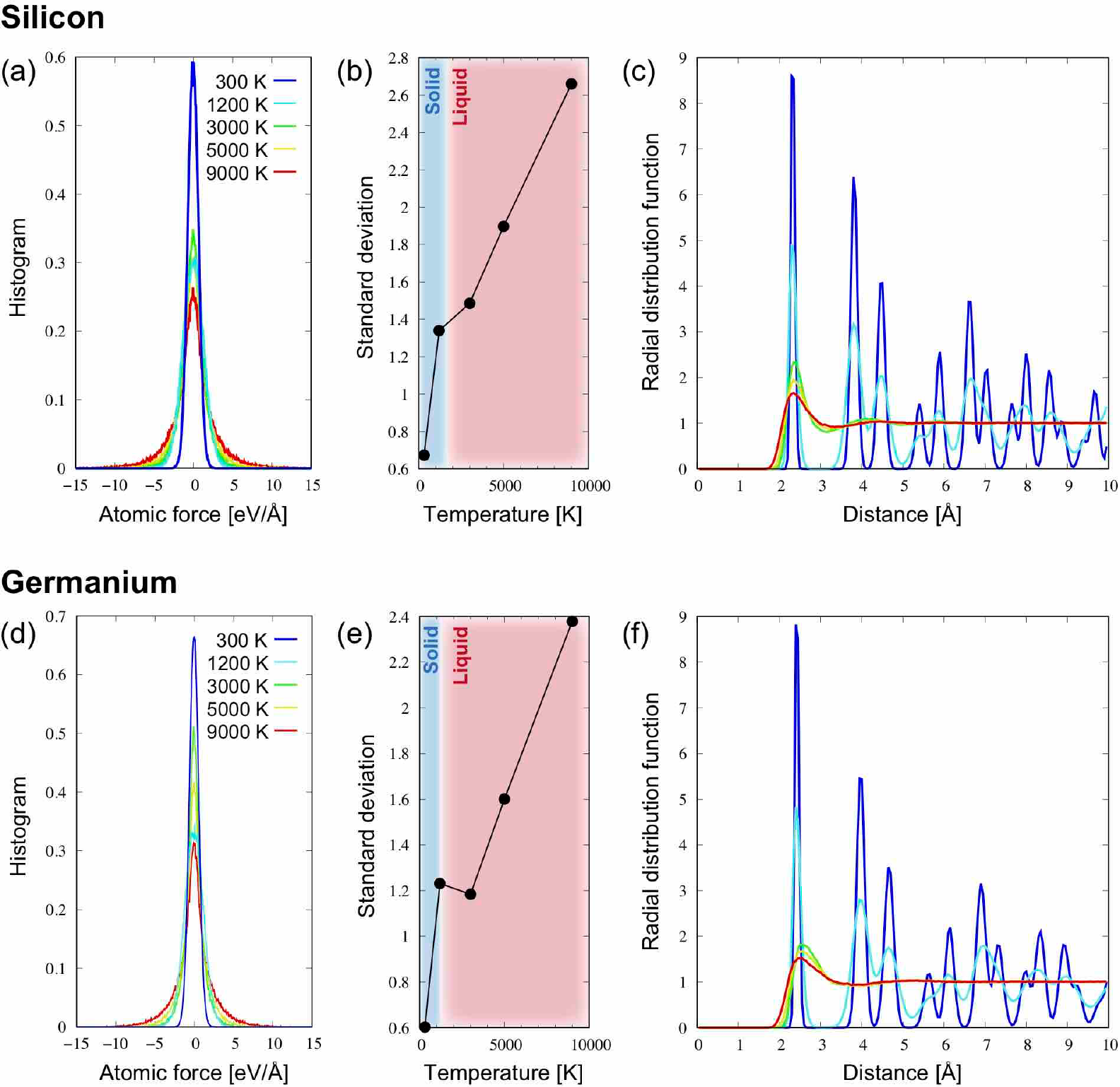} 
\end{center}
\caption{\label{fig:Force}
(Color online) 
(a) Histogram of atomic forces,
(b) temperature dependence of standard deviation $\delta$ of atomic forces,
and (c) radial distribution function in the system consisting of 1000 Si atoms.
(d) Histogram,
(e) standard deviation,
and (f) radial distribution function in the system consisting of 1000 Ge atoms.
}
\end{figure*}

Using the calculation conditions explained in the last section, DFT-MD simulations are performed at 300 K, 1200 K, 3000 K, 5000 K, and 9000 K for both Si and Ge single component systems.
Here, we consider the homogeneous systems which do not have defects, surfaces, and interfaces.
As seen in the following, the simulations at 300 K and 1200 K cases correspond to the solid phase, while those at 3000 K, 5000 K, and 9000 K to the liquid phase for both Si and Ge systems.
In each case, we prepare 2 ps simulation with a time step of 1 fs after the equilibration,
and then 2000 configurations $(u=1,...,2000)$ are obtained at each temperature.

Figures~\ref{fig:Force} (a) and (d) are histograms showing the temperature dependence of atomic forces in the Si and Ge systems, respectively.
Here, to draw a histogram,
we take a random sampling from the configuration $u=1,...,2000$ and atomic index $i=1,...,1000$. The number of sampled points is $10^5$.
Gaussian-like distributions centered on the origin are observed in each histogram.
Furthermore, 
in Figs.~\ref{fig:Force} (b) and (e),
the temperature dependence of the standard deviation $\delta$ of atomic forces (e.g., the width of histograms of Figs.~\ref{fig:Force} (a) and (d)) is plotted.
In the Si system, broadening of atomic forces is monotonically increased against temperature.
On the other hand, in the Ge system, $\delta$ decreases a little
at the boundary between the solid and the liquid states.
Figures~\ref{fig:Force} (c) and (f) are the radial distribution function(RDF):
\begin{eqnarray}
g (r) = \frac{\langle n(r) \rangle}{4 \pi r^2 \Delta r \rho},
\end{eqnarray}
where $\langle n(r) \rangle$ is the average number of atoms in spherical shell within $r$ and $r+\Delta r$.
Here, $\Delta r$ is set as $0.05 \, {\rm \AA}$,
and $\rho$ is the average density of atoms
($\rho= 0.050 \, {\rm \AA^{-3}}$ for Si and $\rho= 0.044 \, {\rm \AA^{-3}}$ for Ge).
Apparently,
shapes of RDF are different between the solid (300 K and 1200 K) and liquid (3000 K, 5000 K, and 9000 K) states.
As increasing the temperature, all peaks are broadened,
and those except for the nearest-neighbor peaks almost disappear in the liquid phase.

%%%%%%%%%%%%%%%%%%
\section{Accuracy of ML forces at each temperature} \label{sec:acurracy}
%%%%%%%%%%%%%%%%%%

%%%%
\subsection{Training models depending on the temperature}

\begin{figure*}[t]
\begin{center}
\includegraphics[trim=0mm 0mm 0mm 0mm ,scale=0.8,angle=0]{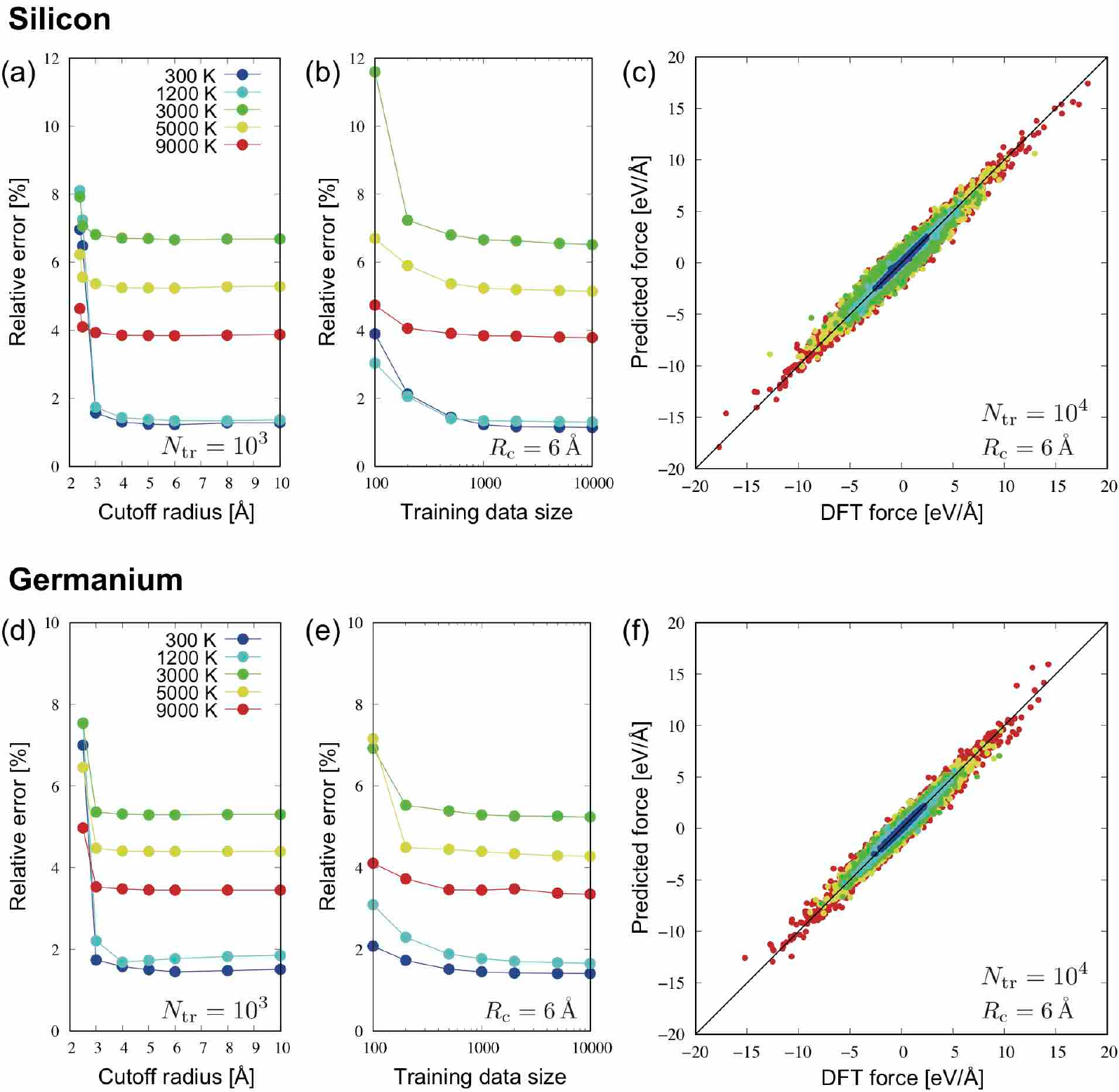} 
\end{center}
\caption{\label{fig:error}
(Color online) 
(a) Cutoff radius ($R_{\rm c}$) dependence of relative error for $N_{\rm tr}=10^3$,
(b) training data size ($N_{\rm tr}$) dependence of relative error for $R_{\rm c} = 6 \, {\rm \AA}$,
(c) parity plot of DFT force and predicted force by ML model trained at each temperature for $N_{\rm tr}=10^4$ and $R_{\rm c} = 6 \, {\rm \AA}$ in the Si system.
(d) Cutoff radius dependence of relative error,
(e) training data dependence of relative error,
(f) parity plot of forces in the Ge system.
}
\end{figure*}

To investigate the accuracy of predicted forces by ML model trained at each temperature,
we first consider the case where the training and the test data are generated at a same temperature.
The data are sampled at random from configuration and atomic indices $u$ and $i$.
The direction $\mathbf{e}$, along which force component is calculated, is also randomly selected for each sample.
Note that the configuration, atom, and the force-component direction are selected independently between training and test data sets.
Hereafter, we express the numbers of data points of training data and test data as $N_{\rm tr}$ and $N_{\rm te}$, respectively.
The accuracy of ML force is evaluated by the mean absolute error (MAE) between the DFT and the ML forces for test data set. 
Furthermore, since the magnitude of the forces is very different depending on the temperature, 
we introduce a relative error defined as MAE/5$\delta$, using the standard deviation $\delta$ in the distribution of forces, as in Ref.\cite{Suzuki-2017}.

Figures~\ref{fig:error} (a) and (d) show a cutoff radius ($R_{\rm c}$) dependence of the relative error, when $N_{\rm tr}=10^3$, $N_{\rm te}= 10^4$ and the fingerprint dimension is $K=100$.
Here, for $\eta_k$, logarithmic grid\cite{Botu-2015} up to $R_{\rm c}$ is adopted.
It is not shown here, but we confirm that the relative error decreases as $K$ increases, 
and $K=100$ is enough to achieve the convergence.
From Figs. ~\ref{fig:error} (a) and (d), we see that the error of ML forces converges very
quickly with respect to $R_{\rm c}$ especially for the liquid state; at 3 - 4 ${\rm \AA}$ for the liquid state
and around 6 ${\rm \AA}$ for the solid state.
Roughly speaking, atoms up to fifth neighbors are included in the range of 6 ${\rm \AA}$
from the target atom (see Figs.~\ref{fig:Force} (c) and (f)). 
For the range of 3 - 4 ${\rm \AA}$, only the nearest and some of the second nearest atoms 
are contained. 
This result indicates that our trained ML model is strongly based on the positions of
the neighbor atoms in a very short range, especially for liquid state.

Next, we consider the number of training data ($N_{\rm tr}$) dependence of relative error for each temperature when $ R_{\rm c} = 6 \, {\rm \AA}$ is used, which is shown in Figs.~\ref{fig:error} (b) and (e).
They show that the relative error does not change largely over $N_{\rm tr}=10^3$, 
and $N_{\rm tr}=10^4$ is enough to achieve the convergence.
Using $N_{\rm tr}=10^4$, 
Figs. ~\ref{fig:error} (c) and (f) show the parity plots of the DFT and ML forces,
and we summarize the value of MAE, relative error, and determination coefficient $R^2$ in Table~\ref{tab:data}.
Although MAE monotonically increases as increasing the temperature, 
relative error and determination coefficient show different behaviors;
they increase in the solid state, while they decrease in the liquid state
when the temperature goes up. 
This result means that the accuracy of ML forces around the phase boundary between the solid and the liquid states is lower than those at other temperatures, in our trained ML model.
However, even in the liquid state,
the relative error is smaller than 6.5\% for the Si system and 5.2\% for the Ge system, respectively.
ML models with such accuracy are useful to perform MD simulations to calculate physical properties of materials.

\begin{table*}[h]
\caption{Mean absolute error (MAE), relative error (MAE/$5\delta$), and determination coefficient ($R^2$) for $N_{\rm tr}=10^4$ and $R_{\rm c} = 6 \, {\rm \AA}$ in the Si system(left table) and Ge system(right table) depending on the temperature.}
\label{tab:data}
\begin{center}
\begin{minipage}[t]{.48\textwidth}
\begin{tabular}{r|ccc}
\hline\hline
Si & MAE & Relative error & $R^2$ \\
\hline
300 K& 0.0386 eV/\AA & 1.1467\% & 0.9947 \\
1200 K& 0.0873 eV/\AA & 1.3074\% & 0.9928 \\
3000 K& 0.4839 eV/\AA & 6.5213\% & 0.8320 \\
5000 K& 0.4852 eV/\AA & 5.1516\% & 0.8944 \\
9000 K& 0.5009 eV/\AA & 3.7856\% & 0.9426 \\
\hline\hline
\end{tabular}
\end{minipage}
\begin{minipage}[t]{.48\textwidth}
\begin{tabular}{r|ccc}
\hline\hline
Ge & MAE & Relative error & $R^2$ \\
\hline
300 K& 0.0423 eV/\AA & 1.4076\% & 0.9921 \\
1200 K& 0.1019 eV/\AA & 1.6564\% & 0.9878 \\
3000 K& 0.3099 eV/\AA & 5.2367\% & 0.8900 \\
5000 K& 0.3421 eV/\AA & 4.2757\% & 0.9254 \\
9000 K& 0.3986 eV/\AA & 3.3514\% & 0.9524 \\
\hline\hline
\end{tabular}
\end{minipage}
\end{center}
\end{table*}

\subsection{Discussion of ML forces in liquid phase}
Here, we discuss the reason why the accuracy of ML forces is better at higher temperatures in the liquid state.
First, we would like to note that we cannot improve the accuracy of ML forces in the liquid state even if we increase the cutoff radius larger than 3 or 4 \AA, as we see in Figs.~\ref{fig:error}(a) and (d).
It means that the ML forces do not change even if we displace the atoms whose distance from the target atom is larger than 3 or 4 \AA. 
If the DFT forces also have this property, this behavior of ML forces would not cause any problems.
To check this aspect, we calculate the difference of atomic forces when the atoms outside of a radius $R_{\rm r}$ from the target atom are displaced by 0.2 \AA, with random directions. 

\begin{figure}[h]
\begin{center}
\includegraphics[trim=0mm 0mm 0mm 0mm ,scale=1,angle=0]{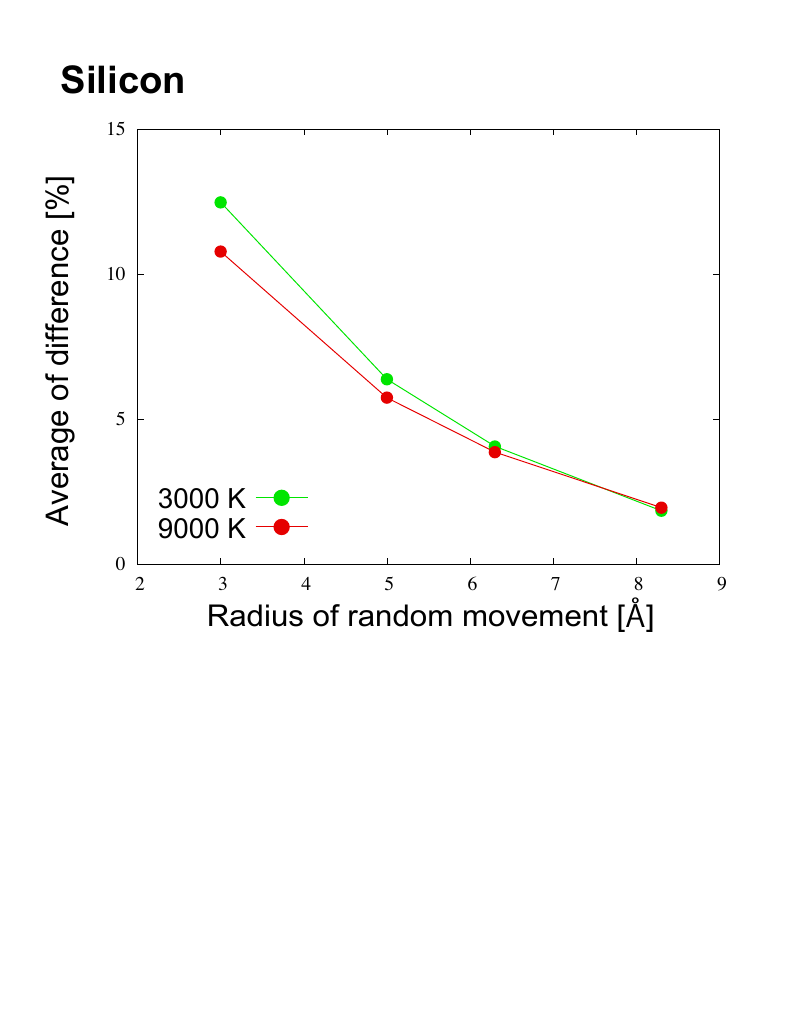} 
\end{center}
\caption{\label{fig:random_force}
(Color online) 
Radius of random movement ($R_{\rm r}$) dependence of the average of normalized difference ($\Delta F_{\rm av} (R_{\rm r})$) defined by Eq.~(\ref{eq:deff_random}) for 3000 K and 9000 K in the Si system.
}
\end{figure}

Figure~\ref{fig:random_force} shows $R_{\rm r}$ dependence of the average of normalized difference 
($\Delta F_{\rm av} (R_{\rm r})$) for 3000 K and 9000 K which is defined as
\begin{eqnarray}
\Delta F_i (R_{\rm r}) = \frac{\left| \mathbf{F}_i^{\rm origin} - \mathbf{F}_i^{\rm random} (R_{\rm r}) \right|}{\left| \mathbf{F}_i^{\rm origin} \right|},
\end{eqnarray}
\begin{eqnarray}
\Delta F_{\rm av} (R_{\rm r}) = \frac{1}{N_{\rm r}} \sum_{i=1}^{N_{\rm r}} \Delta F_i (R_{\rm r}). \label{eq:deff_random}
\end{eqnarray}
Here, $\Delta F_{\rm av} (R_{\rm r})$ is the difference between atomic force in the original configuration 
$\mathbf{F}_i^{\rm origin}$ and the one, $\mathbf{F}_i^{\rm random} (R_{\rm r})$, calculated after 
the outer atoms are displaced.

In the calculation of $\Delta F_{\rm av} (R_{\rm r})$, we use 81 data;
nine atoms are randomly selected from 1000 atoms in the 2000 configurations, and for each case, 
we generate nine different configurations where random movements are performed depending on $R_{\rm r}$.
From Fig. ~\ref{fig:random_force}, we see that the effect of the neighbor atoms whose distance is
larger than $5 \AA$ is not negligible, $\Delta F_{\rm av} (5 \AA)$ is larger than 5\%.
In addition, the force difference at 3000 K is larger than the one at 9000K. 
This implies that the effect of the neighbor atoms in the middle range (3-6 \AA) for the
atomic forces is important especially for the case at 3000 K, and it becomes smaller when 
the temperature increases. It is also consistent with the RDF in Figs.~\ref{fig:Force} (c) and (f).

These considerations suggest that the effect of neighbor atoms in the middle range (3-6 \AA) is 
important to accurately reproduce the DFT forces in the liquid state, especially at low temperatures. 
The ML technique in this work, with the present atomic fingerprint, is difficult to reproduce 
this middle-range behavior, and this is the reason why the present ML forces is more accurate
at higher temperatures in the liquid state.

%%%%%%%%%%%%%%%%%%
\section{Transferability of ML models to temperatures} \label{sec:trans_temp}
%%%%%%%%%%%%%%%%%%

\subsection{Training models in the liquid state}

\begin{figure*}[h]
\begin{center}
\includegraphics[trim=0mm 0mm 0mm 0mm ,scale=0.8,angle=0]{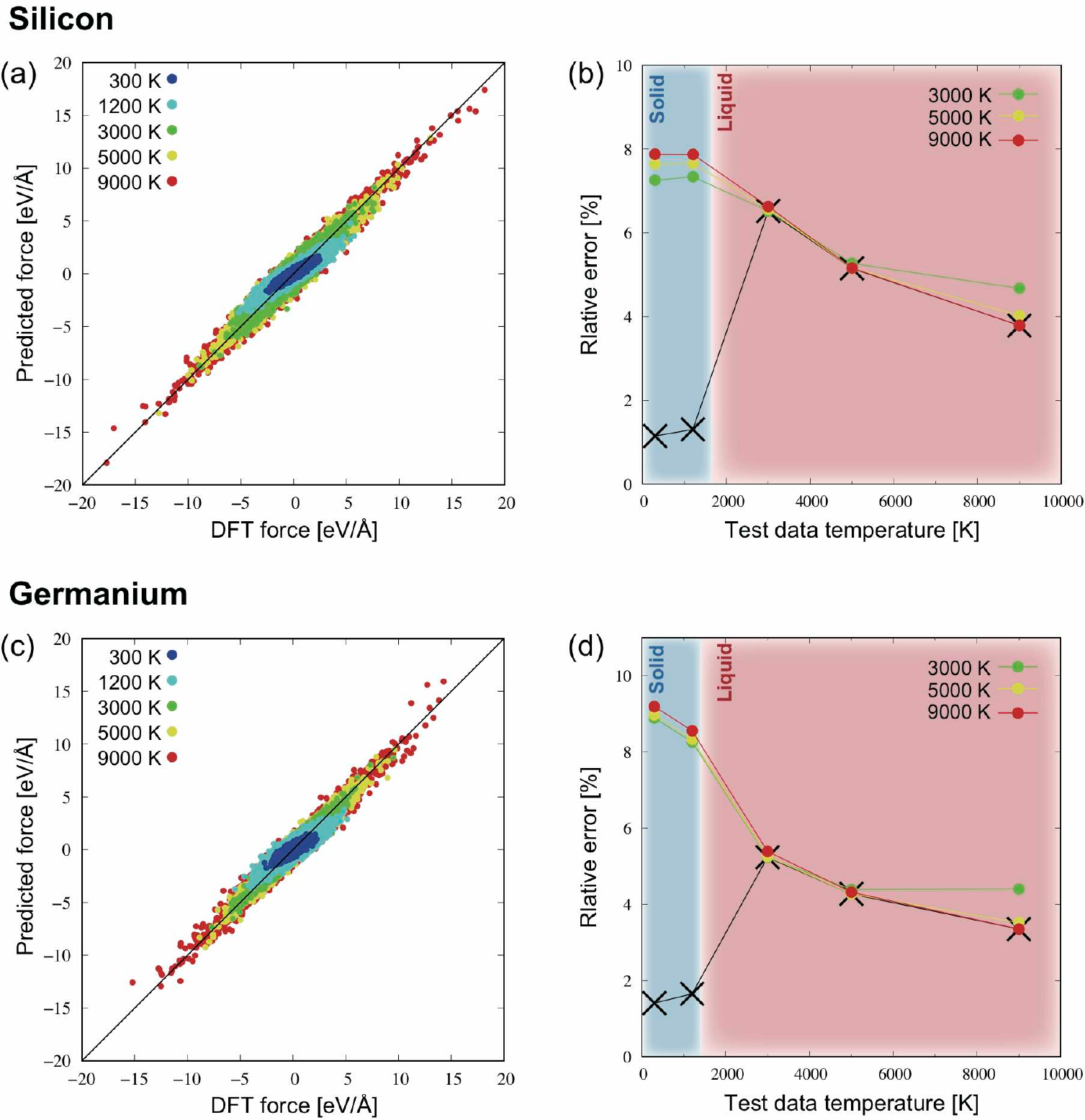} 
\end{center}
\caption{\label{fig:deff}
(Color online) 
(a) Parity plot of the DFT force and predicted force by ML model trained at 9000 K for the Si system.
Each colors indicate the test data temperature dependency.
(b) Test data temperature dependence of relative errors for the Si system.
Green, yellow, and red circle points indicate the results when DFT-MD simulations at 
3000 K, 5000 K, and 9000 K are used as training data, respectively.
Black cross points are the results when the temperatures of training and test data are the same.
(c) Parity plot and (d) relative error for the Ge system.
}
\end{figure*}

To address the transferability of ML model trained in the liquid state to other temperatures, we first apply
the ML model at 9000 K to the predictions of forces at various temperatures.
Figures~\ref{fig:deff} (a) and (c) are the parity plot of the DFT force and predicted one by the ML model 
trained at 9000 K, for test data at various temperatures.
In the liquid state, 
the results are similar to those of the ML forces trained at each temperature
shown in Figs.~\ref{fig:error} (c) and (f).
On the other hand, in the solid state, atomic forces are found to be underestimated in general.
 
We also investigate the transferability of the ML forces trained at different temperatures in the liquid state, as shown in Figs.~\ref{fig:deff} (b) and (d).
Here, green, yellow, and red circle points indicate the relative errors of the ML forces trained at 3000 K, 5000 K, and 9000 K, respectively, for the test data at various temperatures.
We see that the relative error becomes smaller by the increase of temperature, in all cases.
For comparison, the error of the ML forces trained at each temperature is also presented by black crosses in the figure.
In the liquid state,
the errors of the ML forces trained at the corresponding temperature (black crosses) and those evaluated for the ML model trained at different temperatures (colored circles) are close to each other.
In particular, for the trained ML model at 9000 K, the difference of the errors is only 0.11\% for the Si system and 0.16\% for the Ge system.
Thus, we conclude that the transferability of the ML model to different temperatures is high enough in the liquid state when the training is performed at a higher temperature.
The results in Figs.~\ref{fig:deff} (b) and (d) also show that the prediction of forces at high temperature (e.g. 9000 K) is difficult from the training data obtained at lower temperature (e.g. 3000 K).
This is simply because some regions in the fingerprint space for the test data at high temperature are located outside of the space spanned by the training data at low temperatures, 
thus the prediction needs to be made by extrapolation.
On the other hand,
for the solid state, 
the relative error is very large when we use the ML model trained in the liquid state, 
meaning that the ML model trained in the liquid state can not be used for the solid state.
The reason of this low accuracy is discussed in the next subsection.

\subsection{Difference between solid and liquid states from a view point of fingerprint}

Here, we analyze the reason why the accuracy of the ML model trained in the liquid state 
is extremely low for the atomic forces in the solid state. 
There are two possibilities; i) in the fingerprint space, there are no training data 
in the liquid state, which are close to a given test data for the solid state, or 
ii) even though there exists a training data close to a given test data in 
the fingerprint space, the atomic force between these two states may be very different
because there may be important differences of the structure, which cannot 
be properly described by the present fingerprints.
To answer which is the case in the present problem, we evaluate a similarity between atomic 
fingerprints of test and training data.

\begin{figure*}[t]
\begin{center}
\includegraphics[trim=0mm 0mm 0mm 0mm ,scale=0.7,angle=0]{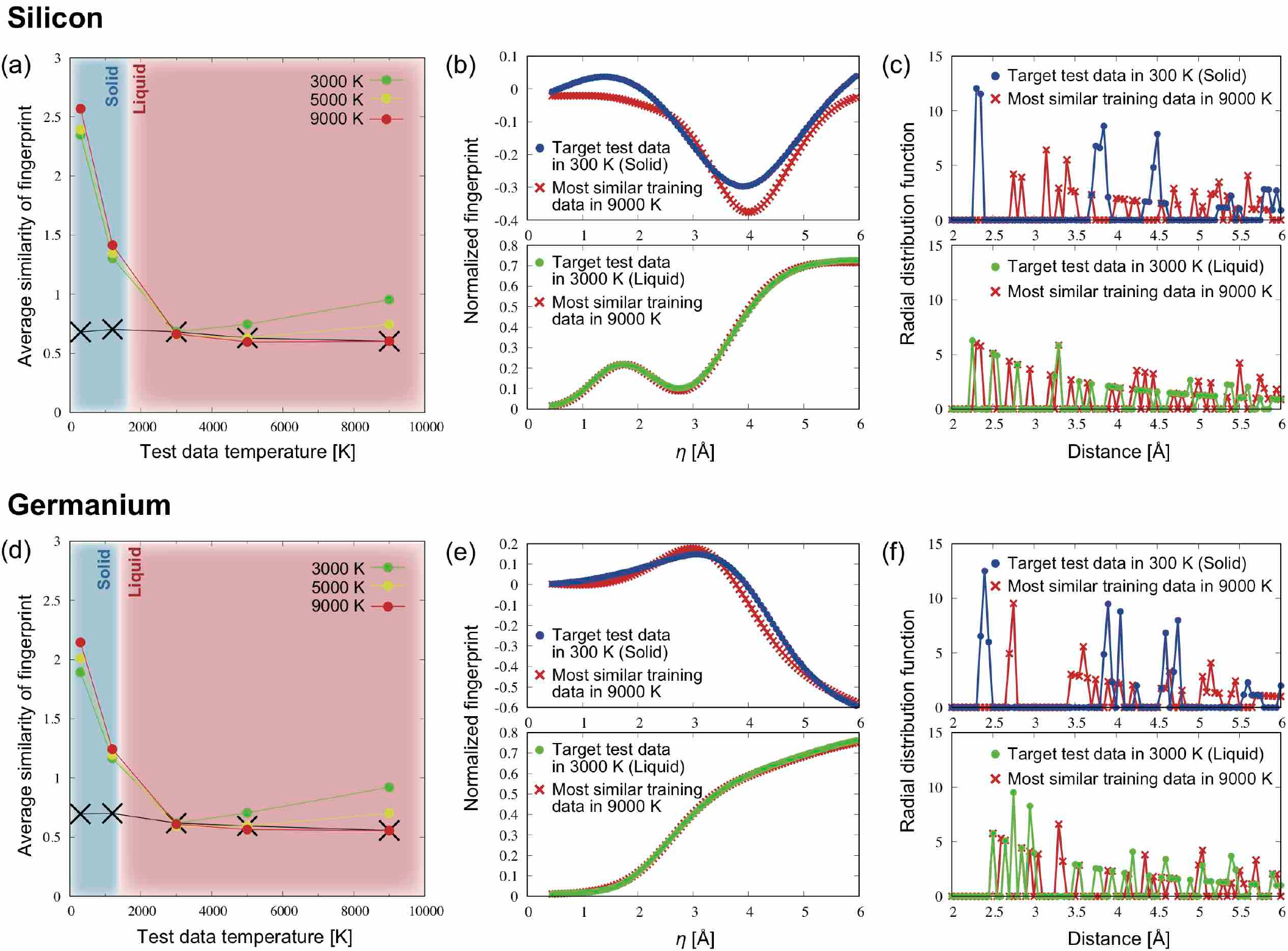} 
\end{center}
\caption{\label{fig:Fin_sim}
(Color online) 
(a) Average similarity of atomic fingerprint defined by Eq.~(\ref{eq:sim_Fin}) for the Si system, at various test data temperature ($T_{\rm te}$).
Green, yellow, and red circle points indicate the results for the training data temperature $T_{\rm tr}=$ 3000 K, 5000 K, and 9000 K, respectively.
Black cross points are the results when $T_{\rm tr}$ is the same temperature with the test data.
(b) Comparison plot of the atomic fingerprints of the target test data and most similar training data in the fingerprint space when the test data temperatures are 300 K and 3000 K, respectively, for the Si system.
(c) Comparison plot of the RDF corresponding to the same data points used in (b) for the Si system.
(d) Average similarity of atomic fingerprint, (e) comparison plot of atomic fingerprints, and (f) comparison plot of the RDFs for the Ge system.
}
\end{figure*}

The similarity of the two data points can be defined by the Euclidean distance of the two points in the atomic fingerprint space. 
For the target test data (index $m$) at temperature $T_{\rm te}$, we express the index of the most similar training data as $m^*$, 
and their similarity between these two data points as $\Delta_m$.
Furthermore, we evaluate the average of the similarity for a set of test data as,
\begin{eqnarray}
\Delta = \frac{1}{N_{\rm te}} \sum_{m=1}^{N_{\rm te}} \Delta_m, \label{eq:sim_Fin}
\end{eqnarray}
where $N_{\rm te}$ is the number of test data and is fixed as $10^4$.

Figures ~\ref{fig:Fin_sim} (a) and (d) show the test data temperature dependence of $\Delta$ for Si and Ge systems, respectively.
Here, we use the different sets of training data at various temperatures ($T_{\rm tr}$) with $N_{\rm tr} =10^4$ at each temperature.
Green, yellow, and red circle points represent the results obtained by the training data at $T_{\rm tr} = 3000$ K, 5000 K, and 9000 K, respectively.
For comparison, we also show the results, by black cross points, when the training data at the same 
temperature ($T_{\rm te}=T_{\rm tr}$) are used.
The results show that the difference between the circle and cross points are not
large in particular for $T_{\rm tr} = 9000$ K, meaning that the average of similarity is small for the test data in the liquid state. 
On the other hand, large differences between the circle and cross points are observed for the solid state.

For more detailed analysis, we check the similarity of each data $\Delta_m$.
Figures ~\ref{fig:Fin_sim} (b) and (e) show the atomic fingerprint for the test data $m$ and training data $m^*$,
which show the smallest value of $\Delta_m$.
Here, we consider the temperature of training data as $T_{\rm tr}=9000$ K, and the comparison of the 
fingerprints for the case at $T_{\rm te}=300$ K and $T_{\rm te}=3000$ K is presented in upper and lower panels, respectively.
Apparently, the difference between the test and training data is small when $T_{\rm te}=3000$ K, while it is large at $T_{\rm te}=300$ K.
The difference in the atomic fingerprint space can be also seen in the RDF for the corresponding data, 
which are shown in Figs.~\ref{fig:Fin_sim} (c) and (f).

From these results, we conclude that the reason for the poor accuracy of the liquid ML model in
the solid state is simply because there are no corresponding training data close to the target test
data in the atomic fingerprint space. 
Note that the ML model trained at high temperature in the solid state (ex. 1200K) is accurate for the 
test data in the solid state at lower temperatures.
It suggests that the present atomic fingerprints is capable to distinguish the local structure 
of the solid and liquid states, and these two states are separated in the atomic fingerprint space.
Following these considerations, it is expected that we can construct a universal ML model 
simply by combining the training data of the solid state and those of the liquid state. 
However, it is found that such mixing models do not work very well.
The detailed results are reported in Appendix.

\begin{figure*}[t]
\begin{center}
\includegraphics[trim=0mm 0mm 0mm 0mm ,scale=0.8,angle=0]{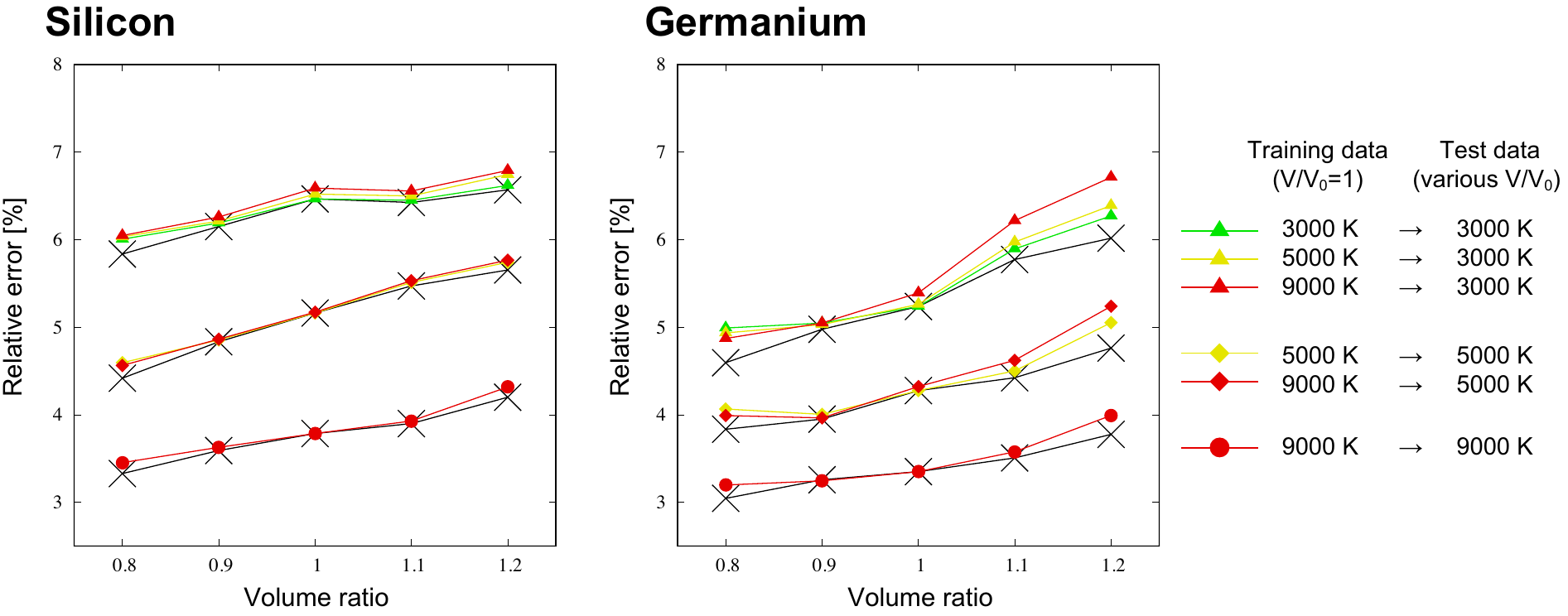} 
\end{center}
\caption{\label{fig:Trans_P}
(Color online) 
Volume ratio ($V/V_0$) dependence of relative errors obtained by the ML model trained in the liquid states for the Si and Ge systems.
Green, yellow, and red points indicate the results by the ML model when DFT-MD simulations at 3000 K, 5000 K, and 9000 K for the standard volume ratio ($V/V_0=1$) are used to prepare the training data, respectively.
Black cross points are the results when DFT-MD simulations for each volume ratio are used to prepare the training data.}
\end{figure*}

%%%%%%%%%%%%%%%%%%
\section{Transferability of ML models to pressure} \label{sec:trans_pre}
%%%%%%%%%%%%%%%%%%

In this section, we study the accuracy of the ML forces in the liquid state, when they are applied
to the test data obtained from the DFT-MD simulations with a different size of the simulation cell. 
So far, we fix the size of the simulation cell, which is a cube with the side length of $L = 27.15 \, {\rm \AA}$ or $L= 28.28 \, {\rm \AA}$ for Si and Ge systems, respectively, as explained in Sec.~\ref{sec:methods}. 
However, when the temperature would change, the volume of the simulation cell should also change. 
In the future targets, there are many systems where some irregular, but important regions feel local stress. 
The accuracy of ML forces in such environments is essential for the future applications with the ML forces. 
In this sense, it is important to investigate the transferability of ML forces to different volumes or
different pressures.

For this purpose, we change the size of the cubic cell from $V_0 (=L^3)$ to $V$; 
$V/V_0 = 0.8$, $0.9$, $1.1$, and $1.2$.
As in the previous sections, the training and test data sets are generated by the random sampling for $u$ and $i$ for each DFT-MD simulation with the simulation time of 2 ps.
Each data set contains $10^4$ data,
and the parameters for atomic fingerprints are $R_{\rm c}=6 \, {\rm \AA}$ and $K=100$.
We first confirm that the deviation of forces is smaller for larger volume (not shown here). 
This is reasonable since the distances between atoms are large and atomic forces are small.

Figure~\ref{fig:Trans_P} illustrates the volume ratio ($V/V_0$) dependence of the relative error in the liquid states.
Cross points in black are the relative errors when the ML forces are constructed by using the training data generated with the same volume and the same temperature used for generating the test data.
We find that the relative error increases as increasing the volume ratio, in most cases.
On the other hand, results of green, yellow, and red points in Fig.~\ref{fig:Trans_P} indicate the volume ratio dependence of the relative errors by the predicted forces for various test data temperatures,
when the ML models were trained at the standard volume $V_0$.
The differences of the relative errors are very small between the predicted forces by the ML model trained at each condition (cross points in black) and those trained at the standard volume $V_0$ (colored points), when $V/V_0$ is $0.9$ and $1.1$; they are smaller than 0.13\% and 0.2\% in the Si and Ge systems, respectively. 
The differences at $V/V_0 = 0.8$ and $1.2$ are larger, but still smaller than 0.22\% in the Si system and 0.4\% in the Ge system. 
From these results, 
we conclude that the transferability of the ML models to pressure is high enough when the change of volume ratio is not very large.

%%%%%%%%%%%%%%%%%%
\section{Discussion and summary} \label{sec:summary}
%%%%%%%%%%%%%%%%%%

In this paper, we investigated a generalization performance of the ML model to predict atomic forces when the training data were sampled from the DFT-MD simulations of the Si and Ge single component systems in the liquid state.
The atomic forces were trained by the Gaussian process regression using the atomic fingerprint defined in Eq.~(\ref{eq:fingerprint}).
We first studied the accuracy of the ML model when the both training and test data were sampled from the DFT-MD simulation at a same temperature. 
The accuracy of the ML model at different temperatures was investigated and it was found that the accuracy is better for higher temperatures in the liquid state, contrary to the temperature dependence in the solid state.
From the detailed analysis of the DFT atomic forces, we found that the effect of neighbor atoms in the middle range (3-6 \AA) is important in the liquid state, especially at low temperatures. 
The atomic fingerprint used in this work is not capable to capture this middle-range behavior, 
and this is the reason why the present ML forces are more accurate at higher temperatures in the liquid state. 

Next, we addressed the transferability of ML models to various temperatures.
We found that in both solid and liquid states, the transferability to temperature is guaranteed,
if the training data are generated at a high temperature in each state.
On the other hand, the accuracy of ML forces was quite low when the ML model trained in the liquid state was applied to the solid state. From the analysis of the similarity between the sets of test and training data, we concluded that the poor accuracy of liquid ML forces for the solid state is simply because there are no corresponding training data close to the test data in the atomic fingerprint space. 
Notice that recently, the transferability to temperature of ML forces by neural network is investigated in Ref.\cite{Kuritz-2018},
and an importance of generalization performance evaluation of each ML model is becoming increasingly.

Moreover, we confirmed that 
the transferability of the ML models to pressure is guaranteed when the change of volume ratio is not very large.
Thus, in the liquid state,
the ML model by training data generated by the DFT-MD simulations under a specific condition,
such as at a high temperature with a standard volume,
can be used to predict accurate forces under various conditions without additional or further DFT-MD simulations.
We believe that the present ML forces have already enough generalization performance for many applications in the computational materials science.

\section*{Acknowledgment}

We thank Teppei Suzuki for the useful discussions at the early stage of the research.
RT was partially supported by Core Research for Evolutional Science and Technology (CREST) (grant number JPMJCR17J2) from Japan Science and Technology Agency (JST)
and the Nippon Sheet Glass Foundation for Materials Science and Engineering. 
JL and TM thank the support by JSPS KAKEHI Grant Number 18H01143 and New Energy and Industrial Technology Development Organization of Japan (NEDO) Grant (P16010).
The computations in the present work were performed on Numerical Materials Simulator at NIMS, 
and the supercomputer at Supercomputer Center, Institute for Solid State Physics, The University of Tokyo.
This work was done as part of the ``Materials Research by Information Integration'' Initiative of the Support Program for Starting Up Innovation Hub, Japan Science and Technology Agency.

\appendix

\section{Mixing of solid and liquid states for training data}

In this appendix, we report the accuracy of a ML model constructed by combining the training data of
two MD simulations in the solid and liquid states.
Here, the training data are sampled from the MD simulations at 1200 K and 9000 K, which are
the highest temperature among the MD simulations in this work for the solid and liquid states, respectively.

\begin{figure}[h]
\begin{center}
\includegraphics[trim=0mm 0mm 0mm 0mm ,scale=0.8,angle=0]{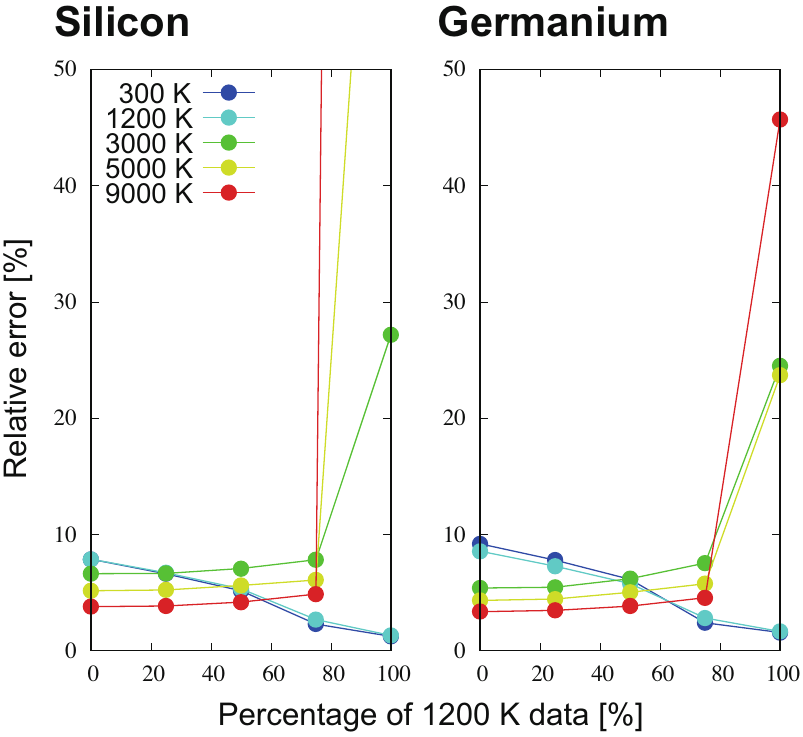} 
\end{center}
\caption{\label{fig:mixing}
(Color online) 
Relative error of ML forces depending on the ratio of 1200 K data in the total training data ($R_{\rm 1200 K}$).
In $R_{\rm 1200 K} =$ 0\%, the training data is only generated at 9000 K.
}
\end{figure}

Figure~\ref{fig:mixing} shows the relative errors of ML forces as a function of the ratio of 1200 K data in the total training data ($R_{\rm 1200 K}$). 
The number of total training data is $N_{\rm tr} = 10^4$, and thus $R_{\rm 1200 K}=$ 50\% means 
the training data includes 5000 data from the solid state and 5000 data from the liquid state.
We use the same test data sets in Secs. 4 and 5 for $N_{\rm te} = 10^4$ to evaluate the relative errors of ML forces.
For the test data in the solid states (300 K and 1200 K),
the relative error monotonically decreases against $R_{\rm 1200 K}$.
On the other hand,
the opposite behavior is observed for the test data in the liquid states (3000 K, 5000 K, and 9000 K).
The minimum relative errors are obtained at $R_{\rm 1200 K} =$ 100\% and 0\% for the solid states and the liquid states, respectively.
The results in Fig. ~\ref{fig:mixing} demonstrate that
it is possible to construct a ML model which shows the error smaller than 7 \% for 
both solid and liquid states, when $R_{\rm 1200 K} =$ 25\% and 50\% for Si case and 50\% for Ge case.
But, they also show that if we can generate the training data in both phases,
it is more accurate to construct two ML models trained in each phase.

%\bibliography{force}
%\bibliographystyle{jpsj}

\end{document}